\documentclass[letterpaper, 10 pt, conference]{ieeeconf}
\IEEEoverridecommandlockouts        
\usepackage{graphicx} 
\usepackage{amssymb}
\usepackage{amsmath,amsfonts}
\usepackage{bm}
\usepackage{listings}
\usepackage{color}
\usepackage{mathtools}
\usepackage[version=4]{mhchem}
\usepackage{comment}

\newtheorem{theo}{Theorem}
\newtheorem{lem}{Lemma}
\newtheorem{defi}{Definition}

\title{\LARGE \bf
Controlling Peak Sharpness in Multimodal Biomolecular Systems via the Chemical Fokker-Planck Equation
}

\author{Taishi Kotsuka and Enoch Yeung
\thanks{*This work was supported by in part by the JSPS Overseas Research Fellowships, an NSF CAREER Award 2240176, the Army Young Investigator Program Award W911NF2010165 and the Institute of Collaborative Biotechnologies/Army Research Office grants W911NF19D0001,  W911NF22F0005, W911NF1920026, and W911NF2320006.  This work was also supported in part by a subcontract awarded by the Pacific Northwest National Laboratory for the Secure Biosystems Design Science Focus Area “Persistence Control of Engineered Functions in Complex Soil Microbiomes" sponsored by the U.S. Department of Energy Office of Biological and Environmental Research.}
\thanks{The authors are with the Biological Control Lab, Department of Mechanical Engineering, University of California at Santa Barbara, Santa Barbara, CA 93106 USA (Taishi Kotsuka: tkotsuka@ucsb.edu), (Enoch Yeung: eyeung@ucsb.edu).}
}

\begin{document}

\maketitle
\thispagestyle{empty}
\pagestyle{empty}

\begin{abstract}
Intracellular biomolecular systems exhibit intrinsic stochasticity due to low molecular copy numbers, leading to multimodal probability distributions that play a crucial role in probabilistic differentiation and cellular decision-making. Controlling the dispersion of multimodal probability distributions in biomolecular systems is critical for regulating stochastic behavior, robustness, and adaptability. However, modifying system parameters to adjust dispersion often affects peak positions, potentially altering a desired phenotype or even fundamental behavior in a genetic pathway. 
In this paper, we establish a theoretical framework that enables independent control of dispersion while preserving peak positions and modality using the Chemical Fokker-Planck Equation (CFPE) and sharpness, a measure of probability concentration around individual peaks. By analyzing the steady-state solution of the CFPE, we derive explicit conditions under which peak sharpness can be tuned monotonically without changing peak positions or modality. We validate our approach through Monte Carlo simulations on a bimodal chemical system, demonstrating effective dispersion control while maintaining structural stability. This framework provides a systematic approach for designing biomolecular systems with tunable stochastic properties, contributing to advancements in synthetic biology and probabilistic cellular regulation.
\end{abstract}

\section{INTRODUCTION}

Intracellular biomolecular systems are known to exhibit stochastic behavior due to the inherently low copy numbers of molecular species, leading to fluctuations in reaction occurrences \cite{Elowitz2002, Raj2008,Paulsson2004}. These stochastic fluctuations can cause variability around steady states and give rise to multimodality in probability distributions. Multimodality in biomolecular systems plays a critical role in cellular differentiation and gene expression variability. It facilitates bistable switching, phenotypic heterogeneity, and population-level robustness, contributing to the functional diversity and resilience of biological systems \cite{Kaern2005, Balazsi2011}. A well known example of behavior driven by multimodality is persister formation in bacterial populations, where a subpopulation enters a dormant state, enhancing survival under antibiotic stress and contributing to antibiotic tolerance \cite{Balaban2004, Niu2024}. From an engineering perspective, designing multimodal systems with precise control over switching mechanisms would enable the development of synthetic biomolecular systems with tailored stochastic behaviors. To achieve this outcome, it is essential to have a fundamental understanding of the distributional characteristics, including the dispersion of each peak, is essential.

\par
\smallskip
In general, the peak positions of a multimodal distribution represent the most probable states or stable configurations of the system, while the dispersion around each peak reflects the intrinsic fluctuations of these states \cite{Risken1989}. Lower dispersion indicate robust states resistant to stochastic fluctuations, whereas higher dispersion suggest increased adaptability to environmental changes through probabilistic state transitions \cite{Friedman2006}. If the dispersion around each peak can be independently controlled, it would enable the design of stochastic biomolecular systems with tunable robustness and adaptability. However, in most cases, modulating the dispersion by altering system parameters simultaneously affects the peak positions and overall modality. Changes in peak positions may result in a shift in steady-state configurations, potentially altering the fundamental behavior of the system. Therefore, it is crucial to identify structural properties of stochastic biomolecular systems that allow control over peak-specific dispersion while preserving peak positions and modality.

\par
\smallskip
In unimodal distributions, variance is commonly used as a measure of dispersion and has been extensively studied in the context of controlling and analyzing stochastic fluctuations. Such analyses typically rely on Chemical Master Equation (CME) modeling, to obtain relationships between variance and system parameters using moment equations or Linear Noise Approximation (LNA) derived from the CME \cite{VanKampen1992,Elf2003,Hayot2004}. 
However, these approaches capture the dispersion across the entire distribution and fail to distinguish the dispersion associated with individual peaks in multimodal systems. A more refined approach is required to characterize peak-specific fluctuations, enabling better control over the stochastic dynamics of multimodal biomolecular systems.

\par
\smallskip
In this paper, we develop a framework for designing multimodal biomolecular systems that can modulate the dispersion of a stationary distribution around individual peaks while preserving peak positions and modality, based on the chemical Fokker-Planck equation (CFPE). Specifically, we first define sharpness in terms of the dispersion of the distribution around individual peaks. We then analyze the parameter dependence of peak positions and sharpness using the steady-state solution of the CFPE, which provides a continuous representation of the probability distribution. Furthermore, we derive structural conditions for biomolecular systems in which sharpness varies monotonically with respect to reaction parameters while maintaining peak positions and modality. Finally, we design a specific biomolecular system based on these conditions and validate that the dispersion around individual peaks can be precisely controlled through Monte Carlo simulations.

\section{Stochastic modeling and problem setting}

\subsection{Chemical master equation}
We consider a univariate biomolecular system consisting $N$ reactions with a molecule $X$, where the $i$-th reaction follows
\begin{equation}
    s_i X \xrightarrow{k_i} (s_i+r_i) X,\label{eq:chemeq}
\end{equation}
where $s_i$ is the number of the molecule $X$ involved in the $i$-th reaction, and $r_i$ is the increased/decreased number of $X$ by the $i$-th reaction. The parameter $k_i$ is the reaction parameter that depends on the reaction rate constant and buffer concentration associated with the $i$-th reaction. While biochemical synthesis typically follows a stepwise manner with $r_i = \pm1$, Eq.~\eqref{eq:chemeq} provides a more flexible modeling framework, capturing cases like multi-cistronic DNA, leading to burst-like synthesis with $r_i > 1$. In general, biomolecular systems are inherently complex, but analyzing a univariate system provides analytical tractability and fundamental insights into stochastic properties, serving as a foundation for extending to more complex multivariate systems.

In a well-mixed system, where molecules react stochastically, the system state follows a continuous-time Markov process. Let $x(t)$ be a discrete random variable representing the molecule count of species $X$ at time $t$. 
Each reaction occurs as a stochastic event, with the probability of the $i$-th reaction in $[t, t+dt]$ proportional to $dt$ and given by the transition rate function $f_i(x)$. Derived from the law of mass action, $f_i(x)$ represents the reaction propensity, quantifying the likelihood of the $i$-th reaction occurring in state $x$. Formally, for the reaction:
\begin{equation}
    f_i(x) = \begin{cases}
        k_i&(s_i=0)\\
        \frac{k_i}{s_i!}\prod^{s_i-1}_{\ell=0}(x(t)-\ell)&(s_i\geq1),
    \end{cases}
\end{equation}
where the product accounts for the combinatorial factors associated with the reaction.

The probability distribution of \( x \), denoted as \( P(x, t) \), evolves according to the Chemical Master Equation (CME):

\begin{equation}
\begin{aligned}
    \frac{\partial P(x,t)}{\partial t} =& \sum^{N}_{i=1}\left[f_i(x-r_i)P(x-r_i,t) - f_i(x)P(x,t) \right].\label{eq:cme}
\end{aligned}
\end{equation}
This equation describes the time evolution of \( P(x, t) \), where the first term represents the probability influx due to reactions increasing the molecule count to \( x \), and the second term accounts for the probability outflux due to reactions decreasing the count.

The total probability must be conserved, which imposes the normalization condition $\sum^{\infty}_0 P(x,t) = 1$. 
As time progresses, the system reaches a stationary distribution $P_s(x)$, which describes the steady-state probability of observing different molecule counts:
\begin{equation}
    P_s(x) \coloneqq \lim_{t\rightarrow \infty}P(x,t)
\end{equation}
This stationary distribution exhibits peaks corresponding to the most probable states. 

\subsection{Problem setting}

\begin{figure}
    \centering
    \includegraphics[width=0.99\linewidth]{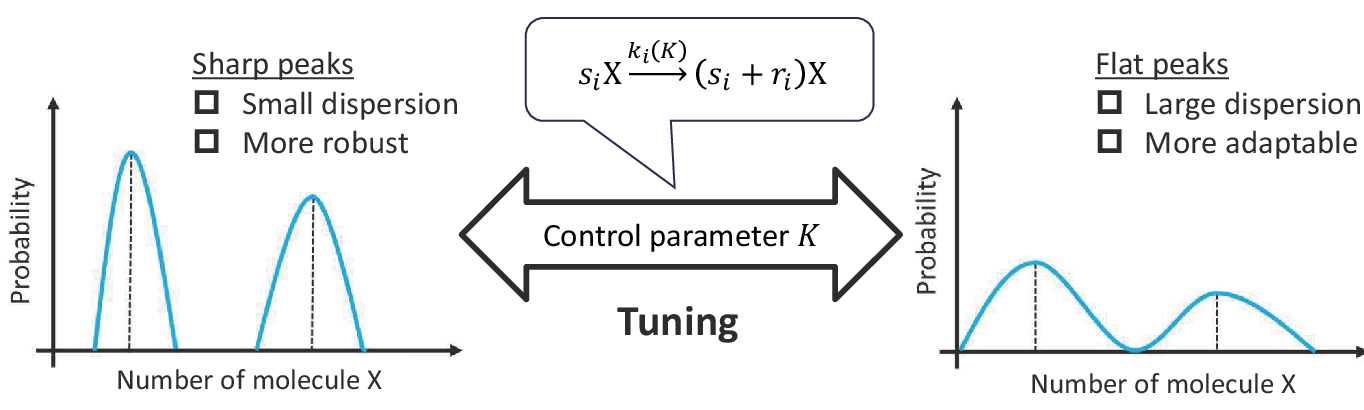}
    \caption{A schematic illustration of a peak sharpness control.}
    \label{fig:schematic}
\end{figure}

A key question is how to control a dispersion of the molecule count around each peak. If the dispersion is low, the system can maintain a desired state despite stochastic fluctuations, whereas the higher dispersion allows probabilistic transitions between states, enabling adaptability to environmental changes as shown in Fig. \ref{fig:schematic}. 
Controlling the dispersion would allow for fine-tuning the system's robustness and adaptability. %as shown in Fig. \ref{fig:schematic}. 
However, modifying system parameters to adjust the dispersion often influences peak positions, which can alter the system's stable states and even compromise its modality. Therefore, we seek a mechanism to adjust the dispersion around these peaks while preserving their positions and the modality. 

To analyze the dispersion around individual peaks of a multimodal distribution, we first introduce a definition of peak positions and valley positions and then characterize a peak sharpness using the probability ratio.

\begin{defi}
    Consider the system (\ref{eq:cme}). We define the peak position $x_{pi}$ and the valley position $x_{vi}$ of the stationary distribution $P_s(x)$ as
    \begin{equation}
    \begin{aligned}
    \{x_{pi} \}_{i=1}^{n} &\coloneqq \left\{x\in\mathbb{R}_{\geq0}|P_s(x)>P_s(x\pm1) \right\},\\
    \{x_{vi} \}_{i=1}^{n-1} &\coloneqq \left\{x\in\mathbb{R}_{>0}|P_s(x)<P_s(x\pm1) \right\}
    \end{aligned}
    \end{equation}
    when $P_s(0)\leq P_s(1)$, and 
    \begin{equation}
    \begin{aligned}
        x_{p1} &\coloneqq 0,\\
        \{x_{pi} \}_{i=2}^{n} &\coloneqq \left\{x\in\mathbb{R}_{>0}|P_s(x)>P_s(x\pm1) \right\},\\
        \{x_{vi} \}_{i=1}^{n-1} &\coloneqq \left\{x\in\mathbb{R}_{>0}|P_s(x)<P_s(x\pm1) \right\}
    \end{aligned}
    \end{equation}
    when $P_s(0)> P_s(1)$. Here, $n$ is the number of peaks, and $x_{v(i-1)}<x_{vi}$ and $x_{p(i-1)}<x_{pi}$.
\end{defi}

\begin{figure}
    \centering
    \includegraphics[width=0.7\linewidth]{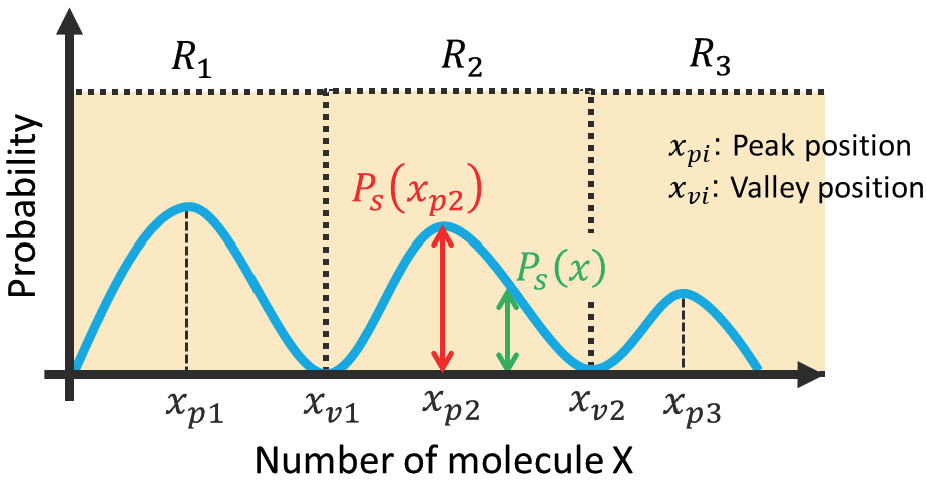}
    \caption{A definition of the peak positions $x_{pi}$, the valley positions $x_{vi}$, and the subregion $R_i$}
    \label{fig:definition}
\end{figure}

The definition is separated in the cases where it has a peak at $x=0$ or not. There must have only one peak between each valley $x_{vi}$ as shown in Fig. \ref{fig:definition}. For analyzing the dispersion of each peak distribution, we separate the region by 
\begin{equation}
R_i \coloneqq
\begin{cases}
[0, x_{v1}), & i = 1, \\
[x_{v(i-1)}, x_{vi}), & 2 \leq i \leq n,
\end{cases}
\end{equation}
where $x_{vn}$ is the supremum of the possible values of $x$. 

\par
\medskip
In a unimodal distribution, variance is generally used as a quantitative measure of the dispersion of a stationary distribution. However, variance cannot effectively assess the dispersion of individual peak distributions since it captures the dispersion across the entire distribution. To analyze the dispersion of individual peak distributions, we define a relative sharpness of the distribution $P_s(x)$ in each subregion $R_i$.

\begin{defi}
    Consider two distinct stationary distributions $P_{s1}(x)$ and $P_{s2}(x)$ that have the same peak $x_{pi}$ and valley positions $x_{vi}$. Then, the distribution $P_{s1}(x)$ is sharper than the distribution $P_{s2}(x)$ in $R_i$ when the probability ratio $\lambda_i(x)$ of $P_{s1}(x)$ is smaller than $\lambda_i(x)$ of $P_{s2}(x)$ for any $x\in R_i$, where
    \begin{equation}
    \begin{aligned}
        \lambda_i(x) &\coloneqq \frac{P_s(x)}{P_s(x_{pi})}, && \mathrm{for}\, x\in R_i.\label{eq:ratio}
    \end{aligned}
\end{equation}
\end{defi}

\smallskip
\par
The sharpness in Definition 2 serves as an indicator similar to variance in quantifying the dispersion of the distribution. While variance measures the overall spread of a distribution, sharpness characterizes the concentration of probability mass around the individual peaks. A higher sharpness implies a more localized distribution with reduced spread, whereas a lower sharpness indicates a broader, more dispersed distribution. Thus, the relative sharpness provides a complementary perspective to variance in assessing the distribution’s shape and dispersion. Therefore, using the probability ratio $\lambda_i(x)$, we elucidate the mechanism that makes the distribution sharper or flatter around the individual peaks while preserving their positions and the modality.

\par
\smallskip
To formalize this design objective, we introduce a control parameter $K$ and consider that some reaction rate constants in the biomolecular system (\ref{eq:cme}) can be tuned by $K$, denoted as $k_i(K)$. The goal is to design a biomolecular system (\ref{eq:cme}) with the control parameter $K$ such that:
\begin{enumerate}
    \item The peak positions $x_{pi}$ and the valley positions $x_{vi}$ remain unchanged by the control parameter $K$.
    \item The sharpness of the stationary distribution $P_s(x)$ in $R_i$ varies monotonically with $K$.
\end{enumerate}
To solve this problem, the parameter dependency of the stationary distribution $P_s(x)$ needs to be analyzed. However, even in the univariate case, obtaining an analytical stationary distribution to the CME is challenging when the state transitions are not limited to nearest neighbors such as a system (\ref{eq:chemeq}). For deriving a closed form of the stationary distribution of multimodal systems, we introduce the Chemical Fokker-Planck equation (CFPE), which is an approximately state-continuous form of the CME. Based on the CFPE, we then analyze the relation between the sharpness of the distribution and the control parameter $K$.

\section{Peak shape characteristics}

In this section, we analyze the dispersion around individual peaks of a multimodal distribution using the CFPE. First, we derive the CFPE as a continuous approximation of the CME. Then, we establish the conditions under which the peak sharpness of a multimodal distribution can be controlled without changing peak positions and modality.

\subsection{Chemical Fokker Planck equation}

We derive the CFPE from the CME for approximately obtaining a stationary distribution of the biomolecular system (\ref{eq:chemeq}). 
By expanding the transition term $f_i(x - r_i) P(x - r_i, t)$ in a Taylor series up to the second order, we obtain
\begin{equation}
    \begin{aligned}
        &f_i(x-r_i)P(x-r_i,t)\\ &= f_i(x)P(x,t) + \sum^{2}_{j=1} \frac{(-1)^j}{j!}r_i^j \frac{\partial^j}{\partial x^j}f_i(x)P(x,t).
    \end{aligned}
\end{equation}
Substituting this expansion into the CME, we obtain the CFPE
\begin{equation}
\frac{\partial P(x, t)}{\partial t} = \frac{\partial}{\partial x} J(x, t),\label{eq:cfpe}
\end{equation}
where the probability flux $J(x, t)$ is given by
\begin{equation}
J(x, t) = A(x) P(x, t) + B(x) \frac{\partial P(x, t)}{\partial x}.
\end{equation}
Here, $A(x)$ is the drift term and $B(x)$ is the diffusion term as 
\begin{align}
    A(x) &= \sum^{N}_{i=1} \left[-r_if_i(x) + \frac{r_i^2}{2}\partial_x f_i(x)\right],\\
    B(x) &= \frac{1}{2}\sum^{N}_{i=1} r_i^2 f_i(x). 
\end{align}

The boundary condition at \( x = 0 \) is determined by the fact that the probability density must vanish for \( x < 0 \). This implies the presence of a reflecting boundary condition at \( x = 0 \), which ensures that the probability flux satisfies $J(0) = 0$. At the steady state, the probability flux must vanish for all \( x \), leading to the equilibrium condition:
\begin{equation}
J(x) = 0, \quad \forall x.\label{eq:flux}
\end{equation}
This formulation provides an effective approximation of the CME, making it more tractable for analysis, particularly in the study of stationary distributions and modality.

By integrating Eq. (\ref{eq:flux}) for $x$, the stationary distribution $P_s(x)$ can be obtained as
    \begin{equation}
        P_s(x) = C\exp{\left(-\int \frac{A(x)}{B(x)}dx \right)},\label{eq:stationary}
    \end{equation}
    where
    \begin{equation}
        C = \left(\int^{\infty}_0 \exp{\left(-\int \frac{A(x)}{B(x)}dx \right)}dx \right)^{-1}.
    \end{equation}
The stationary distribution $P_s(x)$ can be explicitly expressed in terms of $A(x)$ and $B(x)$, which are constructed from the propensity functions $f_i(x)$ of each reaction. This explicit representation allows for the analysis of its dependence on the control parameter $K$.

The gradient of the stationary distribution $P_s(x)$ must be zero at the extrema positions. Using this characteristics, we redefine the peak positions and the valley positions in a continuous-state form as follows. 
\begin{defi}
    Consider the system (\ref{eq:cfpe}). We define the peak position $x_{pi}$ and the valley position $x_{vi}$ of the stationary distribution $P_s(x)$ as
    \begin{equation}
    \begin{aligned}
    \{x_{pi} \}_{i=1}^{n} &\coloneqq \left\{x\in\mathbb{R}_{\geq0}|\partial_x P_s(x)=0\land \partial^2_x P_s(x)<0 \right\},\\
    \{x_{vi} \}_{i=1}^{n-1} &\coloneqq \left\{x\in\mathbb{R}_{>0}|\partial_x P_s(x)=0\land \partial^2_x P_s(x)>0 \right\}
    \end{aligned}
    \end{equation}
    when $\partial_x P_s(0)\geq0$, and 
    \begin{equation}
    \begin{aligned}
        x_{p1} &\coloneqq 0,\\
        \{x_{pi} \}_{i=2}^{n} &\coloneqq \left\{x\in\mathbb{R}_{>0}|\partial_x P_s(x)=0\land \partial^2_x P_s(x)<0 \right\},\\
        \{x_{vi} \}_{i=1}^{n-1} &\coloneqq \left\{x\in\mathbb{R}_{>0}|\partial_x P_s(x)=0\land \partial^2_x P_s(x)>0 \right\}
    \end{aligned}
    \end{equation}
    when $\partial_x P_s(0)<0$, where $x_{v(i-1)}<x_{vi}$ and $x_{p(i-1)}<x_{pi}$.
\end{defi}

\subsection{Conditions for sharpness control without modality change}
A key goal is to adjust peak sharpness while preserving the peak positions and modality. The following lemma shows the condition for ensuring that the positions of the extrema of the stationary distribution $P_s(x)$ remain invariant under changes in the control parameter $K$.

\medskip
\begin{lem}
    Consider the CFPE system (\ref{eq:cfpe}) with the controlled reaction parameters $k_i(K)$ and assume that $A(x)/B(x)$ remains finite for all $x$ in a given finite range. The modality and the extrema positions of the stationary distribution $P_s(x)$ do not depend on the control parameter $K$ if $\partial_KA(x)=0$.
\end{lem}

\par
\smallskip
\begin{proof}
    The derivative of $P_s(x)$ by $x$ is
    \begin{equation}
        \begin{aligned}
            \partial_xP_s(x) &= -\frac{A(x)}{B(x)}P_s(x).
        \end{aligned}
    \end{equation}
    At the extrema, we have $\partial_xP_s(x)=0$ leading to $A(x)=0$ for $P_s(x)\neq0$. The assumption ensures that $P_s(x)\neq0$ for all $x$ in a given finite range. Therefore, the solutions of $A(x)=0$ are the extrema positions of the stationary distribution $P_s(x)$ in a given finite range. If $\partial_K A(x) = \mathbf{0}$ for all $x$, the solutions of $A(x)=0$ do not change by $K$, and thus the extrema positions does not depend on $K$.
    %\qed
\end{proof}

\par
\medskip
This lemma shows that by designing the propensity functions $f_i(x)$ such that $\partial_K A(x)=0$, the peak and valley positions can be preserved regardless of changes in the control parameter $K$. It should be noted that the number of solutions to $A(x) = 0$ determines the modality of the system. We present an example of a biomolecular system satisfying $\partial_K A(x)=0$.

\par
\begin{figure*}
    \centering
    \includegraphics[width=0.85\linewidth]{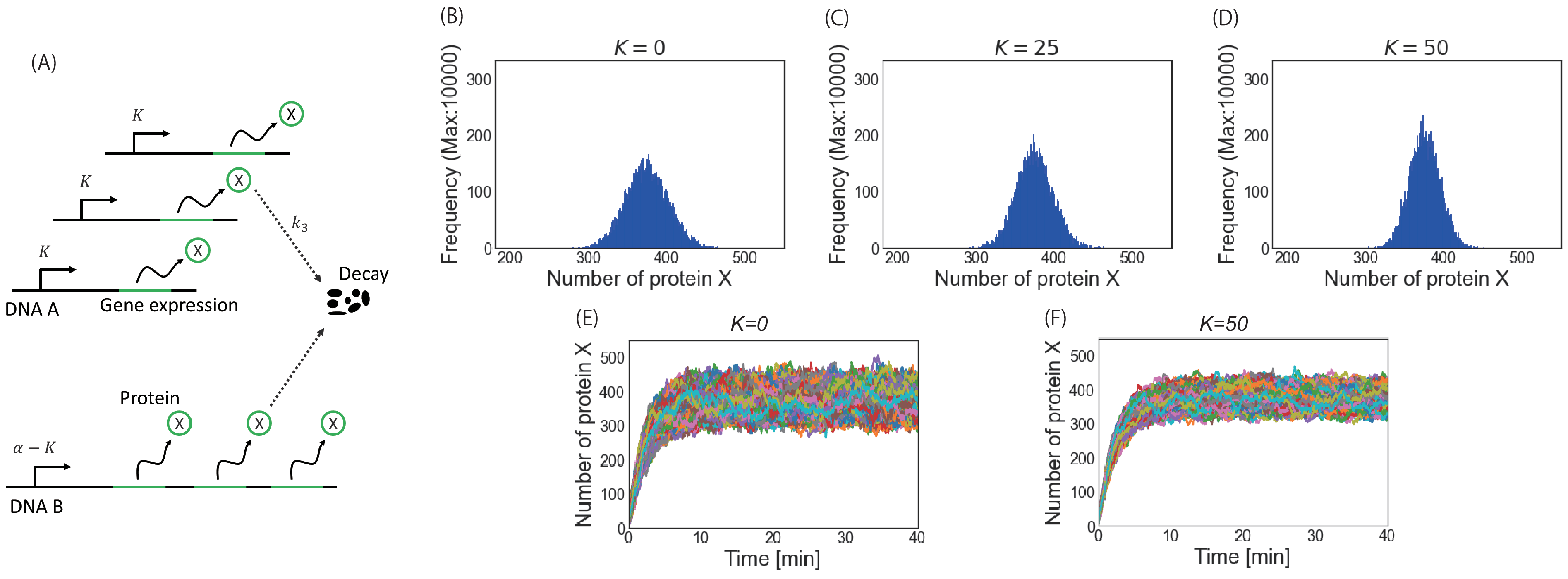}
    \caption{(A) The gene expression of the protein $X$ using two different types of DNA. (B) The distribution of the number of protein $X$ at steady state for 10000 cell for $K=0$, (C) for $K=25$, and (D) for $K=50$. (E) The dynamics of the number of protein $X$ for 1000 cells for $K=0$ and (F) for $K=50$. }
    \label{fig:result_gene}
\end{figure*}

\medskip
\noindent
{\bf Example 1.} Consider a simple biomolecular system in which there are three copies of three copies of a mono-cistronic DNA coding sequence A each encoding the same protein $X$ and one tri-cistronic DNA coding sequence B encoding three copies of the protein X in a single mRNA transcript (Fig. \ref{fig:result_gene}A). The chemical equation is written as
\begin{equation}
\begin{aligned}
    \ce{\emptyset \xrightarrow{k_1} X}, &\quad \ce{\emptyset \xrightarrow{k_2} 3 X}, &\quad \ce{X \xrightarrow{k_3} \emptyset},\label{eq:geneexp}
\end{aligned}
\end{equation}
where $k_1=3K$ and $k_2=\alpha - K$. The parameter $\alpha$ is the production rate of the protein X, $k_3$ is the degradation rate of X. The control parameter $K$ can be tuned by changing the promoter sequence or the copy number of DNA. The function $A(x)$ is
\begin{equation}
    \begin{aligned}
        A(x) &= - 3\alpha + k_3x + \frac{1}{2}k_3,
    \end{aligned}
\end{equation}
which implies that the stationary distribution $P_s(x)$ is unimodal since $A(x)=0$ is the first order equation. It is obvious that $\partial_K A(x)=0$, and thus the change of $K$ does not affect the modality and the extrema positions of the stationary distribution $P_s(x)$. 

\par
\medskip
Next, we present a condition under which the sharpness of the stationary distribution $P_s(x)$ in the subregion $R_i$ varies monotonically with respect to the control parameter $K$ under the condition that the extrema positions are preserved.

\medskip
\begin{theo}
    Consider the system (\ref{eq:cfpe}) with the controlled reaction parameters $k_i(K)$ and assume $\partial_K A(x)=0$. Then, the following holds for all $i$:
    \begin{enumerate}
        \item $\partial_{K_i}\lambda_i(x) \geq 0$ for all $x \in R_i$ if $\partial_K B(x) \geq 0$ for all $x \in R_i$.
        \item $\partial_{K_i}\lambda_i(x) < 0$ for all $x \in R_i$ if $\partial_K B(x) < 0$ for all $x \in R_i$.
    \end{enumerate}
\end{theo}

\medskip
\begin{proof}
    Consider the probability ratio $\lambda_i(x)$. Substituting Eq. (\ref{eq:stationary}) into Eq. (\ref{eq:ratio}), we obtain
\begin{equation}
    \begin{aligned}
        \lambda_i(x) &= \frac{P_s(x)}{P_s(x_{pi})} \\
        &= \exp{\left(\left. -\int \frac{A(x)}{B(x)} \, dx \right|_{x=x_{pi}} + \int \frac{A(x)}{B(x)} \, dx \right)}.
    \end{aligned}
\end{equation}
    Consider the logarithm of the probability ratio $\lambda_i(x)$ as
    \begin{equation}
        \begin{aligned}
            \ln{|\lambda_i(x)|} &= \left[ \int \frac{A(x)}{B(x)} \, dx \right]_{x=x_{pi}} - \int \frac{A(x)}{B(x)} \, dx.\label{eq:lnlambda}
        \end{aligned}
    \end{equation}
    Differentiating both sides with respect to $K$, we obtain
    \begin{equation}
    \begin{aligned}
        G_i(x) &\coloneqq \, \partial_K \ln{|\lambda_i(x)|}\\
        &= \, \left[\int \partial_K \frac{A(x)}{B(x)} \, dx \right]_{x=x_{pi}} - \int \partial_K \frac{A(x)}{B(x)} \, dx,
    \end{aligned}
    \end{equation}
    where $G_i(x_{pi})=0$. Using the condition $\partial_K A(x)=0$, this transforms to
    \begin{equation}
        G_i(x) = \left[\int \frac{A(x)\partial_KB(x)}{B^2(x)} \, dx \right]_{x=x_{pi}} - \int \frac{A(x)\partial_KB(x)}{B^2(x)} dx.
    \end{equation}
    Now consider the derivative of $G_i(x)$ by $x$ as
    \begin{equation}
        \frac{\partial G_i(x)}{\partial x} = \frac{A(x)\partial_KB(x)}{B^2(x)}.
    \end{equation}
    Since $A(x)\leq0$ for $x\in[x_{v(i-1)}, x_{pi})$ and $A(x)\geq0$ for $x\in(x_{pi}, x_{vi})$, the sign of $G_i(s)$ depends on $\partial_KB(x)$. %$G(x)\leq0$ for $x_{v(i-1)}\leq x\leq x_{pi}$ and $G(x)>0$ for $x_{pi}< x< x_{vi}$. 
    Specifically, we have $G_i(x) \geq 0$ when $\partial_KB(x)\geq0$ for all $x\in R_i$ and $G_i(x)< 0$ when $\partial_KB(x)<0$ for all $x\in R_i$. 
    Since $\lambda_i(x)\geq0$ for all $x\in R_i$ and 
    \begin{equation}
        G_i(x) = \partial_K \ln{|\lambda_i(x)|} = \frac{1}{\lambda_i(x)} \partial_K \lambda_i(x),
    \end{equation}
    the sign of $\partial_K\lambda_i(x)$ corresponds to the sign of $G_i(x)$. This leads to the statements of Theorem 1.
\end{proof}

\par
\medskip
This theorem shows that the stationary distribution $P_s(x)$ monotonically flattens as $K$ increases when $\partial_KB(x)\geq0$ while the stationary distribution $P_s(x)$ monotonically sharpens as $K$ increases when $\partial_KB(x)<0$. In summary, the sharpness of the stationary distribution $P_s(x)$ can be monotonically tuned by the control parameter $K$ without changing the modality and the extrema positions by designing the biomolecular system that satisfies $\partial_KA(x)=0$ and $\partial_KB(x)\geq0$ or $\partial_KB(x)<0$.

\section{Numerical example}

To validate the theoretical results derived in the previous sections, we conduct numerical simulations for two biomolecular systems. 

\subsection{Gene expression model}

We first consider the biomolecular system (\ref{eq:geneexp}). For numerical verification, we set the parameter values to $\alpha=50\,\mathrm{molecules\cdot min^{-1}}$, $k_3=0.4\,\mathrm{min^{-1}}$. Analytically, the stationary distribution $P_s(x)$ is a unimodal distribution with the peak at $x_{p1}=374.5$ in the region $R_1=[0,\infty)$. As shown in Example 1, the control parameter $K$ does not affect the modality and the peak position. The function $B(x)$ is
\begin{equation}
    B(x) = \frac{1}{2}(9\alpha + k_3 - 6K),
\end{equation}
and $\partial_KB(x)=-6$. Since $\partial_KB(x)<0$, the stationary distribution $P_s(x)$ in each subregion $R_i$ monotonically sharpens as the control parameter $K$ increases. 

\par
\smallskip
We perform 10,000 Monte Carlo simulations using the Gillespie stochastic simulation algorithm to estimate the stationary distribution $P_s(x)$. The stationary distribution of the number of protein $X$ obtained for different values of the control parameter $K$ are shown in Fig. \ref{fig:result_gene}B - D. The results confirm that increasing $K$ reduces the dispersion of the distribution, sharpening the peak without shifting its position. Additionally, time-series simulations for 10,000 individual trajectories for $K=0$ and $K=50$ shown in Fig. \ref{fig:result_gene}E and F illustrate the higher values of $K$ lead to reduced noise, stabilizing the molecular count over time. The standard deviations are $27.4$ for $K=0$ and $19.6$ for $K=50$, which reduces to $71.25\,\%$. These results suggest that a system generating a small number of molecules with high probability can suppress the noise effect more effectively than that generating a large number of molecules with low probability. Potentially, suppressing the noise effect of the count of the protein $X$ would leads to reduce the noise effect in the downstream system.

\begin{figure*}
    \centering
    \includegraphics[width=0.75\linewidth]{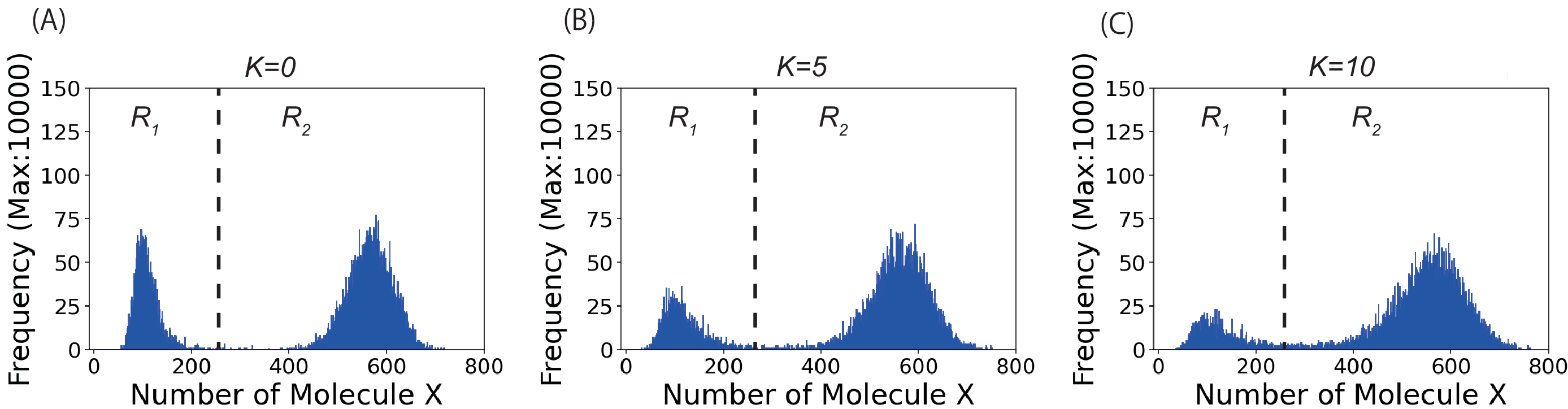}
    \caption{The distribution of the number of molecule $X$ at steady state in the Schlögl system (A) for $K=0$, (B) for $K=5$, and (C) for $K=10$.}
    \label{fig:schlogl}
\end{figure*}

\subsection{Schlögl system}

Next, we consider the Schlögl system, a well-known bistable biochemical system governed by the reactions:
\begin{equation}
\begin{aligned}
    \ce{S_1 <=>[$k_1$][$k_2$] X}, &\,&\,& \ce{S_2 + 2X <=>[$k_3$][$k_{4}$] 3X},
\end{aligned}
\end{equation}
where $S_i$ is the buffer molecule and $k_i$ is the reaction rate. To introduce a control mechanism for peak sharpness, we supplement the system with the following control reactions:
\begin{equation}
\begin{aligned}
    \ce{\emptyset \xrightarrow{k_5} X}, &\quad \ce{X \xrightarrow{k_6} \emptyset}, &\quad \ce{X \xrightarrow{k_7} 2X}.
\end{aligned}
\end{equation}
Here, $k_5=k_6=k_7=K$, ensuring that the newly introduced reactions scale uniformly with $K$. The parameter values are set to $S_1k_1=220$, $k_2=3.5$, $S_2k_3=0.03$, and $k_4=10^{-4}$. The functions $A(x)$ and $B(x)$ for the Schlögl system with the control reactions are computed as 
\begin{equation}
\begin{aligned}
    A(x) =& \frac{k_4}{6}x^3 + \left(\frac{1}{4}k_4 - \frac{1}{2}S_2k_3\right)x^2 + \left(\frac{1}{2}S_2k_3 + k_2\right)x\\
    &- S_1k_1 + \frac{1}{2}k_2,\\
    B(x) =& \frac{1}{12}k_4x^3 + \frac{1}{4}S_2k_3x^2 + \frac{1}{2}(k_2 + 2K)x + \frac{1}{2}S_1 k_1\\ 
    &+ \frac{1}{2}K.
\end{aligned}
\end{equation}
Since $A(x)=0$ is the cubic equation of $x$, the system shows bimodality with these parameter values, where the peak positions are located at $x_{p1}=99.8$ and $x_{p1}=567.6$ with the subregions $R_1=[0,231.1)$ and $R_2=[231.1,\infty)$. It is obvious that $\partial_KA(x)=0$ and $\partial_KB(x)>0$, which suggests the stationary distribution $P_s(x)$ monotonically flattens as $K$ increases without changing the peak positions and modality. 

\par
\smallskip
We conducted 10,000 Monte Carlo simulations to obtain the stationary distribution as shown in Fig. \ref{fig:schlogl}.  In particular, Fig. \ref{fig:schlogl}A shows the bimodal distribution $P_s(x)$ for $K=0$, while Fig. \ref{fig:schlogl}B and C show the distribution for $K=5$ and $K=10$, respectively. The results confirm that the stationary distribution $P_s(x)$ progressively flattens without shifting the peak position as $K$ increases, which corresponds to the analytical results. 

We then focus on the balance of the distributions around the two peaks. The cumulative probability for $x\in R_1$ decreases from $31.19\,\%$ for $K=0$ to $14.75\,\%$ for $K=10$. This result implies that the increase of $K$ further shifted the ratio of low-molecule-number and high-molecule-number states into an imbalanced state similar to persister formation \cite{Blattman2024}, where cells exhibit a bimodal distribution with a dominant normal state and a smaller persister subpopulation. Thus, designing biomolecular systems based on peak sharpness could enable the artificial construction of persister bacterial cells.

\subsection{Robustness analysis of controlled Schlögl system}

In the Schlogl system (29) with the control reactions (30), we initially assume the controller reactions have identical rate constants. To relax this condition, we now consider small perturbations $\delta$ and $\epsilon$ in the rates, denoted as $k_5=K$, $k_6=K+\delta$, and $k_7=K+\epsilon$. With this, the function $A(x)$ becomes
\begin{equation}
\begin{aligned}
    A(x) =& \frac{k_4}{6}x^3 + \left(\frac{1}{4}k_4 - \frac{1}{2}S_2k_3\right)x^2\\
    &+ \left(\frac{1}{2}S_2k_3 + k_2 + \delta - \epsilon \right)x - S_1k_1 + \frac{1}{2}k_2 - \delta, \label{eq:Aperturb}
\end{aligned}
\end{equation}
and $B(x)$ becomes
\begin{equation}
\begin{aligned}
    B(x) =& \frac{1}{12}k_4x^3 + \frac{1}{4}S_2k_3x^2 + \frac{1}{2}(k_2 + 2K + \delta + \epsilon)x\\
    &+ \frac{1}{2}S_1 k_1 + \frac{1}{2}K. \label{eq:Bperturb}
\end{aligned}
\end{equation}
From the first-order and zero-order terms of Eq. (\ref{eq:Aperturb}), we find that the effect of the perturbations on the peak position is negligible when 
\begin{equation}
\begin{aligned}
    |\delta - \epsilon| &<< \frac{1}{2}S_2k_3 + k_2,\\
    |\delta| &<< \left|- S_1k_1 + \frac{1}{2}k_2\right|. 
\end{aligned}    
\end{equation}
Similarly, the first-order term of Eq. (\ref{eq:Bperturb}) suggests that the effect of the perturbations on the peak sharpness can be ignored when
\begin{equation}
    |\delta + \epsilon| << k_2 + 2K.
\end{equation}
These results indicate that perturbations to the three control reaction rates are permissible, provided they remain small relative to the rate constants of the other reactions of the same order. This relaxes the constraints on the applicability of the control reactions.

\section{CONCLUSIONS}
In this paper, we have derived the mathematical conditions for designing univariate biomolecular systems that can modulate the dispersion of a stationary distribution around individual peaks while preserving peak positions and modality based on the CFPE. We have then validated the theoretical results by the Monte Carlo simulations for two specific biomolecular systems. The numerical example has implied that the dispersion control by the proposed method potentially enables the formation of the imbalanced bimodal distribution similar to persister formation. 

In practice, biomolecular systems are often more complex than univariate models. Therefore, our future work will focus on extending this framework to multivariate systems by analyzing the multivariate CME or CFPE. Since obtaining an explicit steady-state solution in such cases is inherently challenging, an effective approach toward this extension would be the marginalization of the multivariate CME \cite{Yeung2013}, which may provide a tractable means of characterizing peak sharpness in higher-dimensional systems.

\addtolength{\textheight}{-12cm}

\bibliographystyle{IEEEtran}


\end{document}